\documentclass[sigplan,nonacm,10pt]{acmart}
\usepackage{hyperref}
\usepackage{empheq}
\usepackage[frozencache,cachedir=.]{minted2}
\setminted{fontsize=\small}
\def\rocqin#1{\mintinline{coq}{#1}}
\usepackage{interval}
\def\itvcc#1#2{\interval{#1}{#2}}
\def\itvoo#1#2{\interval[open]{#1}{#2}}
\def\itvco#1#2{\interval[open right]{#1}{#2}}

\newcommand{\R}{\mathbb{R}}
\usepackage{tikz}
\usetikzlibrary{angles,quotes,arrows.meta}

\title[Formal verification of tilt estimation using the Rocq prover]{Formal verification of tilt estimation\\ using the Rocq prover}



\def\rocq{\textsc{Rocq}}
\def\robotrocq{\textsc{Robot-Rocq}}
\def\mathcomp{\textsc{MathComp}}
\def\analysis{\textsc{MathComp-Analysis}}
\def\minlog{\textsc{Minlog}}

\def\us{\char`\_}
\def\newterm#1{\textsl{#1}}
\def\sect#1{\S~#1}
\def\mydef{\overset{\textrm{def}}{=}}

\begin{document}

\begin{abstract}
The safe operation of a humanoid robot critically relies on
accurate estimation of its vertical orientation, or ``tilt''.
This requires a variety of mathematical tools, including
three-dimensional geometry and differential equations.
To formally verify tilt estimation, we develop a library
for stability analysis in the Rocq prover.
We start by formalizing a theory of ordinary differential equations.
We provide a formalization of the (local) Cauchy-Lipschitz
(a.k.a.\ Picard-Lindel\"of) theorem for existence and uniqueness,
taking advantage of the library support for quotients provided by
the Mathematical Components library.
We extend this formalization with a variant for global existence
and continuous dependence on initial conditions.
Building on these foundations, we develop a theory of Lyapunov
stability that is compatible with an existing formalization of
LaSalle's invariance principle.
We also extend an existing library for the statics of robot
manipulators to support dynamical systems.
Finally, we apply these libraries to the verification of a
state-of-the-art tilt estimation developed for a humanoid robot.
\end{abstract}

\author{Reynald Affeldt}
\affiliation{%
  \institution{National Institute of Advanced Industrial Science and Technology (AIST)}
  \city{Tokyo}
  \country{Japan}
}

\author{Lynda Bentoucha}
\affiliation{%
  \institution{Universit\'e Paris Cit\'e}
  \city{Paris}
  \country{France}
}

\author{Yoshihiro Ishiguro}
\affiliation{%
  \institution{Nagoya University/National Institute of Advanced Industrial Science and Technology (AIST)}
  \city{Nagoya}
  \country{Japan}
}

\author{Holger Thies}
\affiliation{%
  \institution{Kyoto University}
  \city{Kyoto}
  \country{Japan}
}

\maketitle

\section{Towards formal verification of tilt estimation}
\label{sec:introduction}

We identify two motivations for formal verification of humanoid robots.  The first
one is safety. The size and the weight of most humanoid robots are similar to
human beings, which raises safety concerns, for example in case of a fall.
The key to safety is the robot's verticality, i.e., the inclination of gravity
w.r.t.\ to the local vertical axis of the robot, a.k.a.\ \newterm{tilt}.
Its precise estimation allows for detecting a loss of balance and
trigger a dynamic correction of trajectory~\cite[\sect{1}]{benallegue2023tac}
The correctness of this estimation depends on several aspects:
calibration of sensors, correctness of computer programs, and, more
importantly, a validation of the underlying mathematics.
The other motivation is standardization. Though safety standards for mobile robots
are still under development (see, e.g., ISO/WD 25785-1), one reasonably expects
them to incorporate formal proofs, like for IT security with the Common
Criteria~\cite{cc}.

Tilt estimation (and more generally attitude estimation~\cite[\sect{1}]{mahony2008tac})
is typically performed using physical measurements.  For example,
Benallegue et al.\ proposed a tilt estimation scheme for a humanoid
robot using an \newterm{inertial measurement unit} (hereafter, IMU)
and a velocity sensor \cite{benallegue2023tac}.
An IMU consists of a combination of sensors: accelerometer,
gyroscope, and magnetometer.
An accelerometer measures the sum of the gravitational acceleration
and of the linear acceleration, a gyroscope measures the rotation
speed, and a magnetometer measures the direction of the magnetic
field.
The main aspect of the scheme proposed by Benallegue et al.\ is a
differential equation that captures the dynamics of the estimation
error; they studied its properties like the stability of its
solutions.

From the viewpoint of formal verification, the difficulty with tilt
estimation is that it requires a variety of mathematics (3D geometry,
differential equations, stability theory, etc.) that are not yet
available as a consistent library for the \rocq{} prover \cite{rocq}.
The \robotrocq{} library \cite{affeldt2017cpp}, originally based on
the \mathcomp{} library~\cite{mathcompbook}, provides enough 3D
geometry to formalize the forward kinematics problem of robot manipulators
but only small accounts of differential kinematics were added recently.
Rouhling and Cohen have formalized LaSalle's invariance principle
\cite{cohen2017itp} with an application to the inverted pendulum
\cite{rouhling2018cpp} but their work relies on an axiomatization of
ordinary differential equations (ODEs).
A basic theory of ODEs is available in \rocq{} but using a constructive
foundation~\cite{makarov2013itp} that is not directly compatible with
the other libraries above (which depends on classical mathematics
through the \analysis{} library).

In this paper, we explain how we tackle the formal verification
of tilt estimation by combining new developments in ODEs
and stability theory with extensions of existing libraries in the
\rocq{} prover. More precisely, we discuss the following contributions:
\begin{itemize}
\item We formalize a theory of ODEs for the \rocq{} prover, including
  the local Cauchy-Lipschitz theorem and a global variant,
  demonstrating in particular the library support for quotients
  provided by \mathcomp{}.
\item We extend the \robotrocq{} library to handle dynamic systems
  with time-dependent matrices using \analysis{}.
\item We develop a theory of stability including Lyapunov's stability theorem.
\item We provide the first application of dependent type theory to the
  problem of tilt estimation, which includes an application of
  LaSalle's invariance principle as formalized in related work.  In
  particular, our newly developed theory of ODEs allows us to
  discharge every hypothesis of the reused LaSalle formalization.
\end{itemize}

This paper is organized as follows.
In \sect{\ref{sec:formal_model}}, we formalize the physical problem of
tilt estimation and introduce basic elements of our formalization.
In \sect{\ref{sec:formal_diff_equa}}, we present the formalization of
ODEs, including the local existence and uniqueness theorem,
extension to global solutions, and continuous dependence on initial
conditions.  The proof of local existence demonstrates in particular
the use of quotient types in \mathcomp{}.
In \sect{\ref{sec:stability_theory}}, we formalize basic elements of
stability theory.
In \sect{\ref{sec:application_tilt}}, we apply this theory and
previous work on LaSalle's invariance principle to the problem of tilt
estimation.
We review related work in \sect{\ref{sec:related_work}} and conclude
in \sect{\ref{sec:conclusion}}.

\section{Formal model of tilt estimation}
\label{sec:formal_model}

An efficient tilt estimation can be performed using an IMU's accelerometer and gyroscope,
that provide respectively (a)~{\em the linear acceleration} (the sum of the linear
acceleration of the robot and of the acceleration due to gravity) and
(b)~{\em the angular velocity}.
However, this is under the hypothesis that the system has a linear acceleration that
is negligible compared to gravity~\cite[\sect{1}]{hua2016automatica},
which is not the case when a humanoid robot walks.
One way to overcome this limitation is to help the IMU with
independent measurements such as (c)~{\em the linear velocity}~\cite{benallegue2023tac}.
In this section, we deduce from the measurements (a), (b), and (c) the
dynamics of estimation errors in an ODE. This serves as an
introduction to our formalization of the problem of tilt estimation.

\subsection{Modelling of the physical system}
\label{sec:physical_system}

\definecolor{mygreen}{rgb}{0.2, 0.65, 0.3}

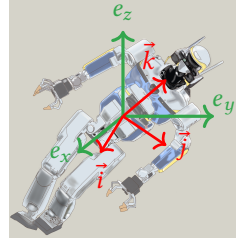
\begin{figure}[h]
\centering
\begin{tikzpicture}
\node[anchor=south west, inner sep=0] (img)
  at (0,0)
  {\includegraphics[width=3cm]{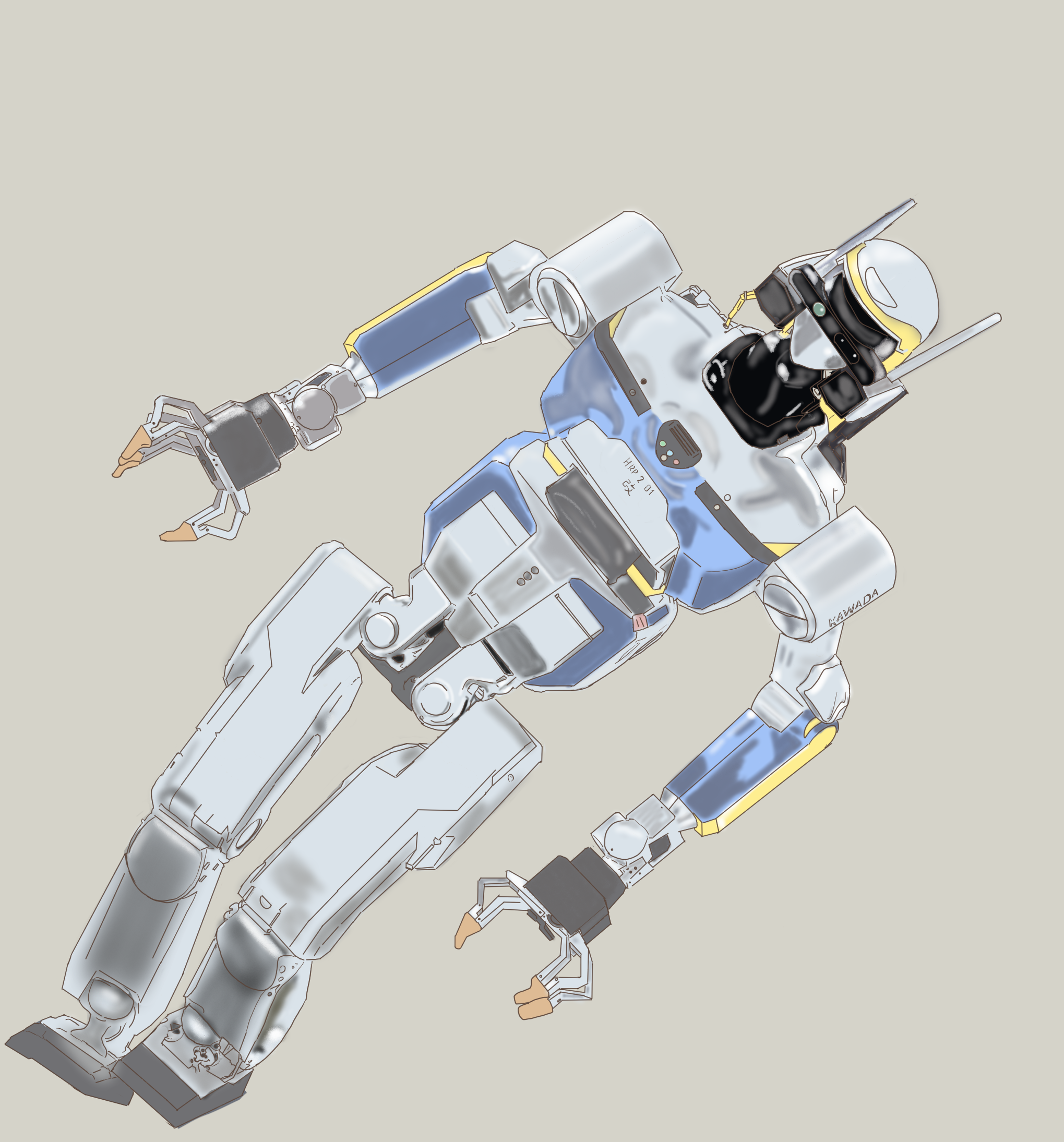}};
\begin{scope}[x={(img.south east)}, y={(img.north west)}]

\coordinate (O) at (0.5,0.5);
\draw[->, very thick, mygreen] (O) -- ++(0.4,0) node[above right,xshift=-1.1ex,yshift=-0.8ex] {$e_y$};
\draw[->, very thick, mygreen] (O) -- ++(0,0.35) node[above] {$e_z$};
\draw[->, very thick, mygreen] (O) -- ++(-0.2,-0.15) node[below left,yshift=1.5ex,xshift=0.5ex] {$e_x$};

\draw[->, very thick, line width=1pt, red] (0.5,0.5) -- (0.69,0.65) node[above left] {$\vec{k}$};
\draw[->, very thick, line width=1pt, red] (0.5,0.5) -- (0.69,0.38) node[right] {$\vec{j}$};
\draw[->, very thick, line width=1pt, red] (0.5,0.5) -- (0.4,0.35) node[below] {$\vec{i}$};

\end{scope}
\end{tikzpicture}
\caption{The world and local frames on the humanoid robot HRP-2 during a loss of balance}
\label{pic:fall}
\end{figure}

\def\framewrt#1#2{{}_{#1}R^{#2}}
\def\phymod{$(\dag)$}
\def\trmx#1{{#1}^{\mathsf{T}}}

Let $W\mydef (e_x,e_y,e_z)$ and $L\mydef (\vec{i}, \vec{j}, \vec{k})$
be respectively the world and the local frames (Fig.~\ref{pic:fall}).
The orientation of frame $L$ w.r.t.\ frame $W$ is
the rotation matrix 
$\framewrt{L}{W}\mydef\begin{bsmallmatrix}
\vec{i}_x & \vec{i}_y & \vec{i}_z \\
\vec{j}_x & \vec{j}_y & \vec{j}_z \\
\vec{k}_x & \vec{k}_y & \vec{k}_z
\end{bsmallmatrix}$
(hereafter, we write $R$ instead of $\framewrt{L}{W}$ to improve readability).
In this matrix, each row corresponds to a basis vector of $L$ but
expressed in the basis $W$ (row convention).
If we knew the position $p$ of the IMU, what the accelerometer
measurements could be written
$\ddot{p}{\trmx{R}} + g_0e_z\trmx{R}$~\phymod,
where $g_0$ represents the gravitational constant (and
$e_z\trmx{R}$ represents $e_z$ but expressed in the local
frame). However, we only know $v=\dot{p}\trmx{R}$ and the angular
velocity~$\omega$.
Benallegue et al.\ therefore model the physical system using the
measurement $y_v$ (linear velocity of the local frame in the world frame
but expressed locally), $y_g$ (angular velocity from the
gyroscope), $y_a$ (measurement of the accelerometer in the local frame) as
follows~\cite[Eqn (1),(2),(3)]{benallegue2023tac}\footnote{
Compared to Benallegue et al., we write $-vS(\omega)$ instead of $S(\omega)v$ because we use row vectors.}:
$$
\begin{array}{rcl}
y_v & = & v \\
y_g & = & \omega \\
y_a & = & - v S(\omega) + \dot{v} + g_0 e_z \trmx{R}.
\end{array}
$$
This is a consequence of the above equation~\phymod{} using the fact
that for a rotation matrix~$R(t)$, we have $\dot{R}(t) = R(t)S(\omega(t))$
with $S(\omega(t))\mydef\begin{bsmallmatrix}
0 & \omega_z(t) & -\omega_y(t) \\
-\omega_z(t) & 0 & \omega_x(t) \\
\omega_y(t) & -\omega_x(t) & 0
\end{bsmallmatrix}$.

\subsection{Ordinary differential equation for the error dynamics}
\label{sec:diff_equa_error}

The system of the previous section can be turned into a differential
equation about the linear velocity and the tilt. Defining
$x_1\mydef v$ and $x_2\mydef e_z \trmx{R}$ (the tilt), we obtain~\cite[Eqn
(11)]{benallegue2023tac}:
\begin{empheq}[left=\empheqlbrace]{align}
\dot{x}_1 & = x_1 S(\omega) + y_a - g_0 x_2 \label{eqn:premiere} \\
\dot{x}_2 & = x_2 S(\omega). \label{eqn:seconde}
\end{empheq}
This inspires Benallegue et al.\ to introduce a \newterm{two-steps first order
  estimator} \cite[\sect{III.B}]{benallegue2023tac}.  Instead of a
direct estimation, they introduce an intermediate estimation
$\hat{x}_2'$ in $\mathbb{R}^3$ from which they derive the estimation
$\hat{x}_2$ (of $x_2$) which is in
$\mathbb{S}^2 \mydef \{x \in \mathbb{R}^3 \mid ||x||^2 = 1\}$ (the
unit sphere centered at the origin) \cite[Eqn
(12)]{benallegue2023tac}:
\begin{equation}
\left\{
\begin{aligned}
\dot{\hat{x}}_1 & = \hat{x}_1 S(y_g) + y_a - g_0\hat{x}_2' \\
\hat{x}_2' & = - \frac{\alpha_1}{g_0}(y_v - \hat{x}_1) \label{eqn12} \\
\dot{\hat{x}}_2 & = \hat{x}_2 S(y_g + \gamma \hat{x}_2' S(\hat{x}_2)).
\end{aligned}
\right.
\end{equation}
Here, $\alpha_1$ and $\gamma$ are two positive real constants.
Benallegue et al.\ further transform the system~\eqref{eqn12} into a
differential equation that captures the dynamics through time of the
estimation error. The estimation errors are defined as
$z_{1}\mydef (x_2-\hat{x}_2')\trmx{R}$ and $z_2\mydef (x_2-\hat{x}_2)\trmx{R}$
and the final differential equation is~\cite[Eqn
(14)]{benallegue2023tac}:
\begin{equation}
\left\{
\begin{aligned}
\label{eqn14}
\dot{z}_{1} & = - \alpha_1z_{1}  \\
\dot{z}_2 & = \gamma (z_2 - z_{1}) S^2(e_z - z_2).
\end{aligned}
\right.
\end{equation}

The goal is to show that the tilt estimate $\hat{x}_2$ converges to the true tilt $x_2$, 
or equivalently that $z_2 \to 0$ for almost all initial conditions.
The proof of this convergence (and even establishing convergence at all) relies on several tools from stability theory and the theory of 
ODEs.
The rest of the paper develops the toolset required for the (first part of the) convergence analysis:
\begin{theorem}[First part of theorem 1 in ~\cite{benallegue2023tac}]
\label{thm:theorem1}
The time-invariant ODE defined by \eqref{eqn14} satisfies the
following: The ODE leaves invariant the set 
$\Upsilon_1 \mydef
\mathbb{R}^3\times \{z\in\mathbb{R}^3\,|\,(e_z - z)\in\mathbb{S}^2\}$,
on which it admits two equilibrium points, namely, $(0,0)$ and $(0,2e_z)$,
and all solutions of \eqref{eqn14} starting in $\Upsilon_1$ converge to one of the two
equilibrium points.
\end{theorem}
Since $\hat{x}_2$ is restricted to have unit length, 
the set $\Upsilon_1$ is the physically meaningful 
state space where the dynamics of the system live.
The theorem therefore shows that starting from a physical meaningful configuration,
the system will remain in such a configuration and converges to one of the two equilibrium points.

\subsection{\rocq{} formalization for the tilt ODE}
\label{sec:tilt_ode}

To introduce formal notations used in the rest of
the paper, let us explain how we formalize Equation \eqref{eqn14}.
Throughout the paper, we work in a context where \rocqin{R} denotes a
numerical type representing the real numbers\footnote{More precisely,
in the formalization, we do not commit to a specific type for real numbers.
Instead, all statements are parameterized by a type \texttt{R : realType} to which
real numbers belong.}.
Let \rocqin{M} be a differentiable, time-dependent rotation matrix. 
We define the tilt as:
\begin{minted}{ssr}
Definition x2 t : 'rV[R]_3 := 'e_2 *m M t.
\end{minted}
In \mathcomp, \rocqin{'rV[R]_n} is the type of row vectors of size \rocqin{n}
with coefficients in \rocqin{R} and matrix multiplication is denoted by~\rocqin{*m}.
The vector \rocqin{'e_2} is the vertical, unit vector of the world frame.
We now assume two constants \rocqin{alpha1} and \rocqin{gamma} and
define the right-hand side of the second equation of Equation~\eqref{eqn14}:
\begin{minted}{ssr}
Definition eqn14b_rhs x1 x2 := 
  gamma *: (x2 - x1) *m \S('e_2 - x2) ^+ 2.
\end{minted}
The notation \rocqin{\S( ... )} is for the skew-symmetric matrix seen
in \sect{\ref{sec:physical_system}} (notation from \cite[\sect{5.1}]{affeldt2017cpp}),
the notation \rocqin{*:} is for multiplication of 
a matrix by a scalar, and \rocqin{^+} denotes exponentiation by a natural number, 
defined by iterated multiplication.

Finally, we formalize Equation \eqref{eqn14} as a function that
receives a vector, splits it into its left and right parts (that is
the purpose of the functions \rocqin{Left} and \rocqin{Right} below),
transforms them individually, and pastes the results back using
\rocqin{row_mx}:
\begin{minted}{ssr}
Definition eqn (dot_z1_z2 : 'rV[R]_6) : 'rV[R]_6 :=
  let dot_z1 := Left dot_z1_z2 in
  let dot_z2 := Right dot_z1_z2 in
  row_mx
    (- alpha1 *: dot_z1)
    (eqn14b_rhs gamma dot_z1 dot_z2).
\end{minted}
We also formally derive the two equations from the physical model in 
lemmas \rocqin{derive_error1_p} and \rocqin{derive_error2_p}.
See \cite{code} for a complete account of the definitions and the
proofs explained in this section.
Hereafter, we refer to the above equation as \rocqin{Tilt.eqn}.

\section{Formalization of ordinary differential equations}
\label{sec:formal_diff_equa}

In this section, we explain an original formalization of the
Cauchy-Lipschitz theorem (to the best of our knowledge, the first
general formalization in the \rocq{} prover, see
\sect{\ref{sec:related_work}}), its continuation to global solutions
under containment in a compact set, and continuous dependence on
initial conditions.
One specificity is the use of quotient types in the local existence
proof to apply the Banach-fixed point theorem.
As far as the application to robotics is concerned, the results
provide the foundations for the subsequent formal verification of
stability analysis.

\subsection{Cauchy problems and the Cauchy-Lipschitz theorem}

A \newterm{Cauchy problem} for an ODE is a triple of a function
$\phi : \R \times \R^n \to \R^n$, an initial time $t_0 \in \R$, and
an initial value $u_0 \in \R^n$. 
We say that $f: \R \to \R^n$ is a solution of the Cauchy problem $(\phi, t_0, u_0)$ on
an interval $I \subseteq \R$ containing $t_0$ if
\begin{itemize}
\item $f(t_0) = u_0$ and
\item for all $t\in I$, $f$ is derivable and $\dot{f}(t)=\phi(t,f(t))$.
\end{itemize}
Hereafter, we refer to the first variable of $\phi$
as the time variable (type $\mathbb{R}$) and the second variable
as the state variable (type $\mathbb{R}^n$).

In general, a Cauchy problem can have no solution, a unique solution, or
uncountably many distinct solutions.
The Cauchy-Lipschitz (or Picard–Lindelöf) theorem is one of the most fundamental theorems in the theory of ODEs;
it establishes the existence and uniqueness of solutions to Cauchy problems. 
\def\cball#1#2{{\mathrm{\overline{B}}}_{#1}(#2)}
The key ingredient used in the theorem is Lipschitz continuity in the state variable. 
We use the following formulation:
\begin{definition}[Lipschitz condition]
Let $|| \cdot ||$ denote a norm on $\mathbb{R}^n$, $a < b$, $u_0 \in \mathbb{R}^n$, and $r > 0$, and let
\[
\cball{r}{u_0} \mydef \{ z \in \mathbb{R}^n \mid \|z - u_0\| \leq r \}
\]
be the closed ball of radius $r$ centered at $u_0$.
A function $\phi: \R \times \R^n \to \R^n$ is said to satisfy a \newterm{Lipschitz condition} in the state variable
on the cylinder $\itvcc{a}{b} \times \cball{r}{u_0}$  
if there exists a constant $L > 0$ such that
\[
\forall x,y\in \cball{r}{u_0},
\forall t\in\itvcc{a}{b}, || \phi(t,x) - \phi(t,y) ||\leq L|| x - y ||.
\]
\end{definition}
In the literature, several equivalent formulations of the Cauchy-Lipschitz theorem coexist, 
differing mainly in the precise specification of solution intervals and endpoint conventions.
In informal mathematical practice, such variations are typically interchangeable, and thus 
often glossed over.
In a formal verification setting, however, the exact formulation must be made explicit and 
we therefore need to carefully specify all the domain requirements, etc.

We choose the following version as the target of our formalization:
\begin{theorem}[Local existence and uniqueness]
\label{thm:cauchy_lipschitz_local}
Let $a < b$, $t_0 \in \itvoo{a}{b}$, $u_0 \in \mathbb{R}^n$, and $r > 0$.
Assume that $\phi : \mathbb{R} \times \mathbb{R}^n \to \mathbb{R}^n$
is continuous in the time variable and satisfies a Lipschitz condition in the state
variable on $\itvcc{a}{b} \times \cball{r}{u_0}$.
Then there exists $\delta > 0$, with $\itvoo{t_0-\delta}{t_0+\delta} \subseteq \itvoo{a}{b}$, 
such that the Cauchy problem $(\phi,t_0,u_0)$ has a unique solution
on $\itvoo{t_0-\delta}{t_0+\delta}$.
\end{theorem}


\subsection{Formal statement of Cauchy problems}

\def\mylim#1#2#3#4{#1 \xrightarrow[ #3 \to #4 ]{} #2}

To formalize \autoref{thm:cauchy_lipschitz_local}, we first introduce
the notions of being a solution of and ODE and of a Cauchy problem in
\analysis.
%
To simplify the presentation of the \rocq{} code, we assume throughout
the rest of this paper a dimension \rocqin{n} and use the notation
\rocqin{U} for the type of row vectors \rocqin{'rV[R]_n} (unless stated otherwise).

We first define a predicate saying that a function~$f$ satisfies the
ODE $\dot f(t) = \phi(t,f(t))$ on an interval:
\begin{minted}{ssr}
Context (phi : R -> U -> U).
Definition sol_is_deriv (A : interval R)
    (f : R -> U) :=
  {in A, forall t, derivable f t 1 /\
    f^`() t = phi t (f t)}.
\end{minted}
In \analysis, derivation is denoted by \rocqin{^`()} or \rocqin{'D_1} and
\rocqin{derivable f t 1} denotes that \rocqin{f} is differentiable at \rocqin{t};
\rocqin{1} represents the direction in which the input varies, 
here this corresponds to the usual one-variable derivative.
The type \rocqin{interval R} denotes the type of intervals over \rocqin{R}.
It can represent bounded and unbounded intervals and records 
whether each finite endpoint is open or closed.
Although we could have used an arbitrary subset of \rocqin{R} for \rocqin{A} in the definition,
intervals are the natural time domains for solutions of ODEs and 
using the interval type simplifies reasoning about endpoints, etc., 
for our later results. 
Following standard ODE terminology, we sometimes call $\phi$ the \newterm{right-hand 
side function} of the ODE.

Our strategy to prove \autoref{thm:cauchy_lipschitz_local} is to first prove
a version whose solution only proceeds forward in time from the initial
value, i.e., for $t \geq t_0$, obtain a solution backward in time
by reversing the time variable, and then combining the two solutions to 
obtain the two-sided version. 
This proof organization is also used in many presentations of the theorem in 
the literature.
For the one sided version, instead of having an additional parameter $t_0$,
 we assume the left endpoint $a$ of the time interval on which $\phi$ is considered 
 to be the initial time. 
 However, in that case we have to be careful about the precise definition of a solution.
In general, we cannot expect two-sided differentiability at $a$ as the function might not even be 
defined on the left of $a$.
As \rocqin{sol_is_deriv} uses differentiability at a point, requiring \rocqin{sol_is_deriv} on the closed interval $\itvcc{a}{b}$ would 
thus be too strong a condition.
On the other hand, \rocqin{sol_is_deriv} on the open interval $\itvoo{a}{b}$ does not suffice, 
as it would not guarantee that the solution is unique.
However, this can be amended by additionally requiring a solution to be continuous 
on the whole interval.\footnote{Alternatively, we could require that the ODE 
holds at finite interval endpoints using only one sided derivatives, 
but the corresponding theory is less developed in \analysis{}. } 
More precisely, we require that solutions are continuous in
the subspace topology defined by the closure of the interval~\rocqin{A}:
\begin{minted}[numbers=left,xleftmargin=3.0ex,escapeinside=88]{ssr}
Definition is_sol (A : interval R) (f : R -> U) :=
  sol_is_deriv A f /\
  {within (closure [set` A]), continuous f}. 8\label{line:within_closure}8
\end{minted}
The \rocqin{within} construction from \analysis{} in
line~\ref{line:within_closure} expresses continuity with respect to
the subspace topology on the given subset.
In particular, for a closed interval $\itvcc{c}{d}$,
\rocqin{{within `[c, d], continuous f}} means that $f$ is continuous in
$\itvoo{c}{d}$, $\mylim{f(x)}{f(c)}{x}{c^+}$, and
$\mylim{f(x)}{f(d)}{x}{d^-}$ .
In \rocqin{is_sol}, line \ref{line:within_closure} amounts to
ordinary continuity in the interior and one-sided continuity at finite
end-points.
This presentation stems from \analysis{} being based on derivability
at a point instead of using left/right derivability.

Next, we formalize the notion of being a solution to a Cauchy problem
by extending the definition \rocqin{is_sol} with an initial condition:
\begin{minted}{ssr}
Definition is_sol_cauchy (a : R) (b : itv_bound R)
    (u0 : U) (f : R -> U) :=
  f a = u0 /\ is_sol phi (Interval (BRight a) b) f.
\end{minted}
Here, the left endpoint of the interval is given by a real number
\rocqin{a} and it is taken to be left-open (this is the meaning of the
constructor \rocqin{BRight} when used at the left-end), while the
right side \rocqin{b} is a generic interval bound and can potentially be $+\infty$.
We further define \rocqin{is_sol_cauchy_oo} as the variant where both 
interval bounds are open and finite.

\subsection{Formal statement of the Cauchy-Lipschitz theorem}
\label{sec:formal_cauchy_lipschitz}


Equipped with the above definitions, we are ready to state the forward
version of the Cauchy-Lipschitz existence theorem.

\def\safedist{{\delta_0}}
\def\safedistfull#1#2#3#4#5#6#7{\safedist(#1,#2,#3,#4,#5,#6,#7)}
\def\dplus#1#2#3#4#5#6#7{\safedistfull{#1}{#2}{#3}{#4}{#5}{#6}{#7}}

We start by defining the size of the existence interval
\rocqin{safe_dist} ($\delta_0$) explicitly. We do not hide it behind an existential
quantifier, which makes it easier to prove additional properties that
are useful in extensions.
It is defined from the constants used in the hypotheses
(\rocqin{phi}, \rocqin{a}, \rocqin{b}, and \rocqin{u0} from the Cauchy problem,
\rocqin{r} and \rocqin{k} from the Lipschitz condition) and an
additional parameter \rocqin{rho}
\cite[Equation (C.4), p.~658]{khalil2002}\footnote{Unless necessary,
we do not display the parameters of $\safedist$ as they can often be
inferred from the context.}:
\[
\safedistfull{\phi}{a}{b}{u_0}{r}{k}{\rho} \mydef \min\left\{b-a,\frac{r}{kr + h}, \frac{\rho}{k} \right\}\] 
with 
\[
h\mydef\sup_{t\in\itvcc{a}{b}}\{\|{\phi(t,u_0)}\|\} \text{ and } \rho <1.
\]
The parameter $\rho$ can be chosen arbitrarily, but it turns out useful
for other parts of the formalization, for example, in the proof of uniqueness later in this section,
 to not replace it by a fixed constant as it allows to control the size of the existence interval.

To avoid repeating common parameters in each \rocq{} definition, we
adopt the convention that unless stated otherwise we are working in a
context with \rocqin{phi} of type \rocqin{R -> U
  -> U}, \rocqin{a} and \rocqin{b} of type \rocqin{R} (with \rocqin{a <
  b} or \rocqin{a <= b}), \rocqin{u0} of type \rocqin{U}, the
parameters of a Lipschitz condition (\rocqin{r : {posnum R}} and
\rocqin{k : R} non-negative or positive), and a parameter \rocqin{rho : {posnum R}}.
The type \rocqin{{posnum R}} is a \mathcomp{} type for positive
numerical values. When \rocqin{x} has type \rocqin{{posnum R}},
\mintinline{ssr}{x
\begin{minted}[numbers=left,xleftmargin=3.0ex,escapeinside=88]{ssr}
Let B := closed_ball u0 r%:num.
Hypothesis cont1 : {in B, forall y,
  {within `[a, b], continuous phi ^~ y}} 8\label{ref:mathcompplaceholder}8.
Hypothesis lip2 : {in `[a, b], forall x,
  k.-lipschitz_B (phi x)}.
Hypothesis rho1 : rho%:num < 1.

Theorem cauchy_lipschitz_ex :
  is_sol_cauchy_oo phi a (a + safe_dist) u0 f. 8\label{line:f}8
\end{minted}
Like \rocqin{safe_dist}, the solution function \rocqin{f} (line
\ref{line:f}) constructed in the proof is not existentially
quantified; it is defined explicitly as some properties of the
construction are required for extensions.
For example, we additionally show that the solution is continuous
within the existence interval and that its image remains in
$\cball{r}{u_0}$.
The construction of \rocqin{f} is described in the next section
(\sect{\ref{sec:about_formal_proof}}).

The \mathcomp{} notation \rocqin{^~} (line
\ref{ref:mathcompplaceholder}) acts like a placeholder.
The \rocqin{lip2} hypothesis means that \rocqin{phi x} satisfies the
Lipschitz condition with Lipschitz constant \rocqin{k} on the set
\rocqin{B}; the Lipschitz notation is from \analysis{} \cite[file
  \texttt{normed\us{}module.v}]{analysis}.

Mathematically, our forward version is a slight generalization of
\autoref{thm:cauchy_lipschitz_local} as the initial time can be on the
boundary of the domain of the right-hand side function.  The original
two-sided version \autoref{thm:cauchy_lipschitz_local} can be recovered 
as a special case, although recovering differentiability at $t_0$ for 
an initial time in the interior is not completely straightforward. 
We define the symmetric version as follows:
\begin{minted}{ssr}
Definition is_sol_cauchy_sym t0 d u0 (f : R -> U):=
  f t0 = u0 /\ sol_is_deriv phi `]t0 - d, t0 + d[ f.
\end{minted}
Note that, contrary to \rocqin{is_sol_cauchy}, no additional
continuity condition is needed here, since differentiability on the
open interval already implies continuity on the whole open interval.  
Then for any $t_0\in\itvoo{a}{b}$, we have:
\begin{minted}{ssr}
Lemma cauchy_lipschitz_sym :
  is_sol_cauchy_sym phi t0 safe_dist_sym u0
    cauchy_lipschitz_f_sym.
\end{minted}
The size \rocqin{safe_dist_sym} is defined using the safe distance of
the forward Cauchy-Lipschitz theorem.
As for the function \rocqin{cauchy_lipschitz_f_sym}, it is defined by pasting two
solution functions of the forward Cauchy-Lipschitz theorem.
The precise construction is explained along the formal proof in
\sect{\ref{sec:gen_sym_case}} below.

Finally, uniqueness is established as an additional theorem:
\begin{minted}{ssr}
Theorem cauchy_lipschitz_unique g :
  is_sol_cauchy_oo phi a (a + safe_dist) u0 g ->
  {in `[a, a + safe_dist], f =1 g}.
\end{minted}

\subsection{About the formal proof of \autoref{thm:cauchy_lipschitz_local}}
\label{sec:about_formal_proof}

The proof essentially follows the classical argument, reducing the
problem to the study of the fixed point of the Picard operator
$P(f)(t) = u_0 + \int_a^t \phi(x,f(x)) dx$.  The main ingredients
are an application of the Banach fixed-point theorem to show existence
and uniqueness of a fixed point, and the Fundamental Theorem of
Calculus to show that this fixed point is indeed a solution to the
differential equation.
Both theorems are already provided by \analysis{}
\cite[file \rocqin{normed_module.v}]{analysis}
\cite[\sect{10.1}]{affeldt2024itp}, but to apply them in our case we need
a few generalizations.

Before all, we extend integration in \analysis{} (which is only
defined for real-valued and extended real-valued
functions~\cite{affeldt2023jar}) to vector-valued functions.
Concretely, we provide a component-wise definition of integration for
vector-valued functions (in finite dimension):
\begin{minted}{ssr}
Definition rowRintegral (f : T -> U) : U :=
  \row_i (\int[mu]_(x in D) (f x) 0 i).
\end{minted}
In this definition, \rocqin{T} is some measurable type and \rocqin{D} is a set.
The notation \rocqin{\row_i} is to aggregate elements as a row vector;
\rocqin{(f x) 0 i} represents the \rocqin{i}th coordinate of a vector
\rocqin{f x}, because matrices (and vectors in particular) are in
essence formalized as functions with two arguments.
A full generalization to Banach spaces using the Bochner integral is
the topic of ongoing work, but beyond the scope of the current paper. 

\subsubsection{Quotients for continuous functions}
\label{sec:quot_contSeg}

An important step in the formalization is to define an
appropriate Banach space of functions on which we can apply the Banach
fixed-point theorem to show that the Picard operator has a
fixed-point.
More precisely, the functions $f$ passed to the Picard operator should
be equipped with a structure of a complete normed module.

The canonical choice in this setting is the space of continuous
functions over an interval with the supremum norm.
\analysis{} already has a type \rocqin{continuousFunType} for
functions defined on a subspace of a topological space packaged with a
proof that the function is continuous (with respect to the subspace
topology \cite[\sect{5.1}]{affeldt2024itp}) and that its values are
contained in a specified codomain.
We specialize this type to the type of continuous functions over
an interval $\itvcc{a}{b}$ to some normed module\footnote{To define
the Picard operator, we use \rocqin{U := 'rV[R]_n} for some \rocqin{n},
which is a normed module.} $U$ in \rocq{} as
\begin{minted}{ssr}
continuousFunType `[a, b] [set: U]
\end{minted}
and these functions are
shown to form an Abelian group (\rocqin{zmodType}) and a left module
(\rocqin{lmodType}).
The function
$f \mapsto \sup_{x \in \itvcc{a}{b}} \left\{ | f(x) | \right\}$
defines a seminorm over functions of this type \cite[Definition
\texttt{pre\us{}infty\us{}norm}]{code}.

However, \rocqin{continuousFunType} does not change the underlying
function type which in our case is $\mathbb{R} \to U$.
Consequently, taking the supremum over $\itvcc{a}{b}$ does not extend
to a full norm as functions agreeing on $\itvcc{a}{b}$ can differ
outside the interval.
Changing the domain type to, for example, an interval subtype solves
this issue, but it would complicate the formalization as most of the
existing theory on differentiation and integration in \analysis{} is
developed for total functions and would need to be adapted.
Instead, \mathcomp{} offers a convenient solution via quotients
\cite{cohen2013itp}, allowing us to identify functions that are equal
on $\itvcc{a}{b}$.
Unlike, e.g., Lean, which provides quotient types as a primitive
of the underlying type theory together with a quotient soundness axiom,
Rocq does not come with built-in quotient types.
Instead, \mathcomp{} implements the quotient construction
entirely within Rocq's existing type theory, by providing an abstraction
layer based on canonical projections and a choice of
representatives.
This allows us to reason about
equivalence classes using ordinary equality while retaining total
functions as representatives, and hence to reuse the existing
differentiation and integration theory without modifying its domain
types.
For our purpose, we define \rocqin{@quot_contSeg R a b U},
the type of quotients of functions from the interval
\rocqin{`[a, b]} of real numbers to some normed module \rocqin{U}
(\rocq{} notation: \rocqin{`C[a, b]}).
On the quotient type, point-wise constructions that apply only to the
interval remain valid by replacing \rocqin{f : `C[a, b]}
with a canonical representative \rocqin{repr f}.
Moreover, we show that the seminorm defined above is a proper norm and
\rocqin{`C[a, b]} constitutes a complete normed module.

\subsubsection{Overview of the proof for the forward Cauchy-Lipschitz theorem}
\label{sec:fwd_cauchy_lipschitz_proof}

We first explain the proof for the forward Cauchy-Lipschitz
theorem (\rocqin{cauchy_lipschitz_ex} in \sect{\ref{sec:formal_cauchy_lipschitz}}).
Let $\phi$ be a function for which the hypotheses \rocqin{cont1} and \rocqin{lip2}
of the previous section (\sect{\ref{sec:formal_cauchy_lipschitz}}) hold.
For the Picard operator to be defined on some $g$ of type
\rocqin{`C[a, b]}, we need to make sure that the inside of the
integral is integrable, which is the case (thanks to continuity)
whenever $g(t)$ stays in $\cball{r}{u_0}$.
We extend this to a total operator \rocqin{picard_fun} that for
functions $g$ of type \rocqin{R -> U} returns the function defined by
the Picard operator whenever $g$ satisfies this property and the
constant $0$ function otherwise.
Using the safety distance \rocqin{safe_dist} from the previous section
we can show that \rocqin{picard_fun} maps \rocqin{`C[a, a + safe_dist]}
into itself and that it is a contraction mapping on the subset
\rocqin{img_cball} defined by
$\left\{ f \,|\, f\left(\itvcc{a}{a+\safedist}\right)\ \subseteq{} \cball{r}{u_0} \right\}$:
\begin{minted}{ssr}
Definition img_cball : set `C[a, a + safe_dist] :=
  [set f | f @` `[a, a + safe_dist] `<=` B].
\end{minted}
Here, \rocqin{`<=`} is a notion for subset; recall that \rocqin{B}
corresponds to $\cball{r}{u_0}$
(\sect{\ref{sec:formal_cauchy_lipschitz}}).
Since moreover \rocqin{img_cball} is closed and non-empty (it contains
at least the constant function $u_0$), we can apply the Banach
fixed-point theorem to get a unique fixed point \rocqin{picard_fix} of
type \rocqin{`C[a, a + safe_dist]} for the Picard operator.
Finally, by the Fundamental Theorem of Calculus, a function is a
solution to the ODE iff it is a solution to the integral equation,
which completes the proof of existence by choosing \rocqin{f}
to be the representative \rocqin{repr picard_fix}.

Although the fixed-point theorem also states that the fixed-point is unique, 
the proof of \rocqin{cauchy_lipschitz_unique} is a bit more involved.
The fixed-point theorem gives uniqueness only among the functions 
contained in \rocqin{img_cball}, which a priori  does not
exclude other solutions whose image leaves the ball.
To remove this restriction, assume we are given any other solution 
\rocqin{g} on $\itvcc{a}{a+\safedist}$. We use the fact that every 
solution is continuous, and therefore, since both solutions have the 
same initial value, they remain in the required ball for some 
sufficiently short time. By varying the parameter \rocqin{rho}, 
we can adjust the length of $\safedist$, to make the existence interval
short enough to guarantee that both solutions satisfy the 
integral equation and belong to \rocqin{img_cball}.
In fact, since this is needed later, we make sure that the 
solutions are contained in the smaller ball $\cball{r/2}{u_0}$, 
and therefore belong to \rocqin{img_cball}.
On this shorter interval, both solutions are 
equal to the unique fixed-point of the Picard operator \cite[Lemma \rocqin{initial_solution_unique}]{code}.
Finally, we extend this equality to the whole interval $\itvcc{a}{a+\safedist}$. 
For this, consider the set 
$E \mydef
 \left\{t\in\itvcc{a}{a+\safedist}
 \,\middle|\,
 \forall s\in\itvcc{a}{t},\ f(s)=g(s)\right\}$.
Let $s = \sup E$ and assume $s < a + \safedist$.
As $E$ is closed, $s \in E$. 
Thus, the above argument lets us 
extend the uniqueness slightly beyond $s$, contradicting the fact that $s$ 
is the supremum of $E$.
Thus $s=a+\safedist$, and the two solutions agree on the whole interval.

\subsubsection{Generalization to the symmetric case (\autoref{thm:cauchy_lipschitz_local})}
\label{sec:gen_sym_case}

Let us now complete the proof \autoref{thm:cauchy_lipschitz_local}, i.e., the 
case where $t_0 \in \itvoo{a}{b}$ and the time interval for the solution extends 
symmetrically in both directions.
First, we define the function
$\phi^{-}(t, x) \mydef -\phi(-t, x)$ and show that the continuity and
Lipschitz conditions are not violated by this transformation.
We further show that if $f_0^{-}$ is a solution to the ODE
$\dot{f}_0^{-}(t) = \phi^{-}(t, f_0^{-}(t))$ over $\itvoo{-b}{-a}$, then
$f^{-}(t) \mydef f_0^{-}(-t)$ is a solution to the original ODE over
$\itvoo{a}{b}$.
Thus, by applying the forward theorem we can get a solution $f^{-}$
on  \rocqin{]t0 - safe_dist_neg, t0[}, where \rocqin{safe_dist_neg}
is the safe distance for the ODE $\phi^{-}$) and, similarly, a solution $f^{+}$ over
\rocqin{]t0, t0 + safe_dist_pos[} such that $f^{-}(t_0) = f^{-}(t_0) = u_0$.
However, we cannot simply patch them together as from the theorem 
we do not get differentiability at $t_0$. 
Instead we choose a distance \rocqin{safe_dist_sym} strictly smaller 
than both distances\footnote{Precisely, \texttt{safe\us{}dist\us{}sym} is defined as the minimum of
$b-t$, $\safedistfull{\phi}{t_0}{b}{u_0}{r/8}{k}{\rho}$, and
$\safedistfull{\phi^-}{-t_0}{-a}{u_0}{r/8}{k}{\rho}$.} and consider the patch of the two 
functions as the solution to a new ODE starting at 
\rocqin{t0 - safe_dist_sym} with initial value $u_1 \mydef f^{-}(t_0- \texttt{safe\_dist\_sym})$.
By the right choice of \rocqin{safe_dist_sym} we can make sure that this new
solution stays in a small enough ball centered at $u_1$ and thus can
be transformed to the integral equation corresponding to the
definition of the Picard operator which in turn proves that it is a
valid solution to the ODE on the whole interval.

\subsection{Extension to global solutions}
\label{sec:global_cauchy_lipschitz}

The local Cauchy-Lipschitz theorem
(\sect{\ref{sec:formal_cauchy_lipschitz}}) only guarantees the
existence of a solution on a (usually very small) neighborhood of the
initial time. It therefore fails to deal with solutions defined for
all non-negative time, which is however what is required to perform
stability analysis of the long-term behavior of a system.

There are several extensions of the basic existence and uniqueness
theorem that provide conditions under which a local solution can be
extended to a global solution.
Here, by global solution, we generally mean a solution which is valid for
all time $t \geq a$.
Global existence and uniqueness can be proved under the hypothesis
that there exists a global Lipschitz constant \cite[Theorem
  3.2]{khalil2002}. This is however a restrictive condition, not met
by typical models of physical systems, which on the contrary typically
have locally Lipschitz functions.

We therefore focus on the continuation result under more assumptions
about the solution of the system, essentially that the solution
remains in a compact region \cite[Theorem 3.3]{khalil2002}. This
condition naturally arises in the stability analysis of robotic
systems.
Precisely, we have the following assumptions:
\begin{minted}{ssr}
Context (K : set U).
Hypothesis compactK : compact K.
Hypothesis u0K : u0 \in K.
Hypothesis phi_continuous :
  forall x, continuous (phi ^~ x).
Hypothesis phi_locally_lipschitz : forall b, a < b ->
  forall x, exists r k : {posnum R},
    {in `[a, b], forall t,
      k%:num.-lipschitz_(closed_ball x r%:num)
        (phi t)}.
Hypothesis solutions_in_K : forall b sol,
  is_sol_cauchy_oo phi a b u0 sol ->
  sol @` `[a, b[ `<=` K.
\end{minted}
That is, we assume that \rocqin{K} is a compact subset of \rocqin{U}, 
the initial value \rocqin{u0} is contained in \rocqin{K},
and \rocqin{phi} is continuous in the time variable.
We further assume that on any finite interval \rocqin{`[a,b]}, 
\rocqin{phi} is locally Lipschitz in the 
state variable, uniformly in time, meaning that for every state
$x$, there are a radius \rocqin{r} and a Lipschitz constant \rocqin{k} that
apply to \rocqin{phi t} for every $t \in \itvcc{a}{b}$.
Finally, \rocqin{solutions_in_K} states that every solution starting from
\rocqin{u0} at time \rocqin{a} remains in \rocqin{K} throughout its
interval of existence.
As in the local case, we state the global extension theorem by 
explicitly selecting the solution \rocqin{global_sol} and prove
\begin{minted}{ssr}
Theorem compact_is_sol_cauchy_infty :
  is_sol_cauchy phi a +oo%O u0 global_sol.
\end{minted}
Thus, \rocqin{global_sol} starts from \rocqin{u0},
is continuous on $\itvco{a}{\infty}$, 
and satisfies the ODE for every $t>a$.
The proof follows the standard textbook continuation method.
We first derive joint continuity of
\rocqin{phi} in both variables on each set \rocqin{`[a, c] `*` K}.
We then consider the set of right endpoints up to which a solution
exists, which is nonempty due to the local existence theorem.
To construct a solution up to the supremum of these endpoints,
we select solutions on finite intervals and use uniqueness to 
prove that they agree when they overlap.
We then show that if the supremum was finite, we could extend
the solution beyond it, contradicting the fact that it is the supremum.
The main additional steps are the explicit
construction of the local solutions.

We additionally prove that the solution has a
right derivative \rocqin{phi a (global_sol a)} at the initial time:
\begin{minted}{ssr}
Lemma compact_sol_right_derivable : 
    (h^-1 *: (global_sol (a + h) - global_sol a)) 
      @[h --> 0^'+] --> phi a (global_sol a).
\end{minted}
This property is needed later, when we connect our notion of solution to the 
one used in the formalization of LaSalle's invariance principle \sect{\ref{sec:convergence_equilibrium_points}}.
The proof is mostly an application of the mean value theorem to \rocqin{global_sol}.

Finally, uniqueness of the solution on \rocqin{[a, +oo[} follows directly by applying the local uniqueness result to every finite subinterval:
\begin{minted}{ssr}
Lemma global_solution_unique f g:  
  is_sol_cauchy phi a +oo%O u0 f ->
  is_sol_cauchy phi a +oo%O u0 g ->
  {in `[a, +oo[%R, f =1 g}.
\end{minted}

\subsection{Continuous dependence on initial conditions}
\label{sec:continuity_initial}

For the later stability results 
we additionally need that the
solution of an ODE continuously depends on the initial condition.
For that purpose, we formalize a theorem that compares two solutions~$y$
and~$z$ whose initial values~$u_0$ and~$v_0$ (respectively) may differ.
It quantifies the distance between the solution
functions in terms of the distance between the initial values\footnote{This is a specialization of \cite[Theorem 3.4]{khalil2002} that
compares two solutions whose right-hand side functions may furthermore differ
by some perturbation function and that upper-bounds the distance between
the solution functions also in term of this perturbation.
Since we only need continuous dependence on the initial value, we only
prove the special case where there is no perturbation.}:
$\forall t\in\itvcc{a}{b}, |y(t)-z(t)|\leq|u_0-v_0|e^{k(t - a)}$.

More precisely, we assume an interval $[a,b]$ and a closed ball \rocqin{B}
around one of the initial values where the right-hand side \rocqin{phi} is 
continuous in the time variable and Lipschitz continuous in the state variable.
\begin{minted}{ssr}
Context (v0 : U).
Let B : set U := closed_ball u0 r%:num. 
Hypothesis lip2: 
  {in `[a, b], forall t, k%:num.-lipschitz_B (phi t)}.
Hypothesis cont1: {in B, forall y, 
  {within `[a, b], continuous phi ^~ y}}.
\end{minted}
We then consider two solutions with possibly different initial values, 
which both remain in \rocqin{B}.
\begin{minted}{ssr}
Variables y z : R -> U.
Hypothesis soly : is_sol_cauchy_oo phi a b u0 y.
Hypothesis solz : is_sol_cauchy_oo phi a b v0 z.
Hypothesis By : y @` `[a, b] `<=` B.
Hypothesis Bz : z @` `[a, b] `<=` B.
\end{minted}
The distance between the two solutions can then be bounded in terms of the
distance between the initial values:
\begin{minted}{ssr}
Theorem continuous_dependence t : 
  t \in `[a, b] ->
  `|y t - z t| <= `|u0 - v0| * expR (k%:num * (t - a)).
\end{minted}
In particular, on every fixed bounded time interval, solutions starting from nearby 
initial values remain close to each other.

\section{Towards a library for formalization of stability problems}
\label{sec:stability_theory}

In this section, we discuss the formalization of a theory of stability
to prove \autoref{thm:theorem1}.

\subsection{Difference between the general theory of ODEs and stability theory}

Stability theory is most commonly studied for autonomous systems,
i.e., systems that do not explicitly depend on the time variable.
Such systems play a central role in robotics and physics because, in
general, the dynamics is only determined by the current state and does
not depend explicitly on time, and that is usually easier to analyze.
Furthermore, unlike the Cauchy-Lipschitz theorem, where the focus is
on a solution starting from a single initial value, stability theory
typically studies the behavior of a family of solutions starting from
a set of initial values.
In this section we therefore assume that the systems we consider are autonomous,
that is the right-hand side function of the ODE is a 
function \rocqin{phi : U -> U} that does not depend explicitly on time.
All results in \sect{\ref{sec:formal_diff_equa}} can be applied to such systems 
simply by introducing a trivial dependency on time, i.e., by replacing \rocqin{phi}
with the function \rocqin{fun=> phi : R -> U -> U}, which ignores the time argument. 
This replacement does not change the differential equation or its solutions.
Moreover, the conditions required in the previous section become much simpler.
The hypothesis \rocqin{cont1} holds trivially, since the right-hand side is 
constant in time, and \rocqin{lip2} reduces to ordinary Lipschitz continuity.

\subsection{Definitions: Equilibrium points, stability, and reachable set}
\label{sec:equilibrium_point}

An \newterm{equilibrium point}
\cite[p.~112, \sect{4.1}]{khalil2002} is a point \rocqin{x} in which is left invariant by the ODE.  
In other words the
constant function returning \rocqin{x} (denoted by \rocqin{cst x}) is a
solution of the ODE \rocqin{phi}:
\begin{minted}{ssr}
Definition is_equilibrium_point (x : U) :=
  sol_is_deriv (fun=> phi) `[0, +oo[ (cst x).
\end{minted}

Lyapunov stability is concerned with equilibrium points, which in
simple terms means that all solutions starting close enough to
equilibrium points remain close for all time.
More precisely, a point $x$ is said to be \newterm{Lyapunov stable}
\cite[Definition 4.1]{khalil2002} when one can always confine a
solution to a neighborhood of $x$ by choosing an appropriate initial
neighborhood.
As with the tilt example (\sect{\ref{sec:formal_model}}), when
studying stability, it sometimes makes sense to constrain the system
to a subset instead of considering the dynamics of the whole space,
e.g., to exclude states that are physically impossible when considering
the neighborhoods for stability.
We therefore formulate stability relatively to a 
set of admissible states \rocqin{Init}.
All stability notions in this section quantify only over solutions 
starting from a point in init, allowing stability to be proven 
relative to a chosen subspace rather than the whole space $\R^n$.
We define stability formally as follows:
\begin{minted}{ssr}
Context (phi : U -> U) (Init : set U).
Definition is_stable_at (x : U) :=
  forall eps, eps > 0 ->
    exists d, d > 0 /\ forall f D,
      f 0 \in Init ->
      is_sol_cauchy_oo (fun=> phi) 0 D (f 0) f ->
      `| f 0 - x | < d ->
      forall t, t \in `[0, D[ -> `| f t - x | < eps.
\end{minted}
There are two aspects in which our definition differs from many
standard textbook presentations (including \cite{khalil2002}).
First, we do not explicitly assume that a stable point is an
equilibrium point.  However, this is an immediate consequence of the
definition, as if a solution starting at a state \rocqin{x}
contained in \rocqin{Init} exists, \rocqin{is_stable_at x} means that a solution
starting at \rocqin{x} must remain constant for all future time \cite[Lemma \rocqin{stable_equilibrium}]{code}.

Second, existence of the solution for all non-negative time is often
assumed implicitly (e.g., \cite[Definition 4.1]{khalil2002},
\cite[\sect{I.4}]{rouche1977}).  However, most notions can also be
stated relatively to the maximal interval of existence, which is the
parameter \rocqin{D} in the above definition.  Note that we quantify
over all possible solutions and all existence intervals, and thus
whenever a global solution exists and is unique, the definition
becomes equivalent to the textbook one.  While this slightly
complicates the notation, it can be useful especially in a
formalization context, as global existence often can not be shown a
priori but only derived later as a consequence of stability (see,
e.g., \cite[Footnote 1, \sect{4.1}, p.~113]{khalil2002}).

Last, we formally define the notion of reachable set called for by
\autoref{thm:theorem1} as the set of states \rocqin{x} that are
attainable by a solution \rocqin{f}:
\begin{minted}{ssr}
Definition reachable_set Init : set U :=
  [set x | exists f D, f 0 \in Init /\
    is_sol_cauchy_oo (fun=> phi) 0 D (f 0) f /\
    exists t, t \in `[0, D[ /\ x = f t].
\end{minted}

\subsection{Lyapunov functions}
\label{sec:lyapunov_functions}

The stability problem is typically solved by exhibiting a \newterm{Lyapunov
  function}.  A \newterm{Lyapunov candidate function} \rocqin{V} is
a function that is $0$ at a point (usually, an equilibrium point) and
positive elsewhere:
\begin{minted}{ssr}
Definition is_Lyapunov_candidate
    {R} {U : normedModType R} V (A : set U) (x : U) :=
  x \in A /\ V x = 0 /\
    forall z, z \in A -> z != x -> V z > 0.
\end{minted}
To become a \newterm{Lyapunov function}, its \newterm{derivative along
  a solution} \cite[p.~114, \sect{4.1}]{khalil2002} must be
non-positive.
One defines the derivative of a function \rocqin{V} ``along a
function \rocqin{f}'' as:
\begin{minted}{ssr}
Definition derive_along (V : U -> R) (f : R -> U) t := 
  (jacobian1 V (f t))^T *d 'D_1 f t.
\end{minted}
The identifier \rocqin{jacobian1} (from \analysis{}) denotes the
Jacobian matrix, i.e., the partial derivatives matrix of a
vector-valued function.  In the particular case of a derivative along
a solution, we introduce the notation \rocqin{'D~(f) V}. 
This relation allows us to establish the properties of this derivative
by using theories already present in \analysis{} to show that a
candidate Lyapunov function is indeed a Lyapunov function.

\subsection{Lyapunov's stability theorem}
\label{sec:lyapunov_stability_thm}

\newterm{Lyapunov's stability theorem}~\cite[Theorem 4.1]{khalil2002}
shows that a point $x$ is a stable equilibrium point if one
can exhibit a candidate Lyapunov function for this point that
moreover happens to have a non-positive derivative:
\begin{minted}[numbers=left,xleftmargin=3.0ex,escapeinside=88]{ssr}
Context (phi : U -> U) (A Init : set U)
  (V : U -> R).

Hypothesis openA : open A.
Hypothesis Vdiff : forall t, differentiable V t.
Hypothesis V'_le0 : forall D f,
  f 0 \in Init ->
  is_sol_cauchy_oo (fun=> phi) 0 D (f 0) f ->
  forall t, t \in `]0, D[ -> 'D~(f) V t <= 0.

Theorem Lyapunov_stability :
  is_Lyapunov_candidate V A `<=` 8\label{line:subset}8
  is_stable_at phi Init.
\end{minted}
Here, \rocqin{A} is the open set on which \rocqin{V} is positive definite, 
and \rocqin{Init} is the set of admissible initial states 
used in the definition of stability.

The formal proof is carried out first by assuming that point of
interest is $0$ and then it is generalized to any point by a change of
variables. 
The case for the point $0$ uses standard properties of real analysis:
properties of open sets (the set \rocqin{A} is assumed to be open), of
continuous functions, the extreme value theorem, the intermediate
value theorem, etc.
These properties were all available in \analysis{}, beside a minor
generalization of the extreme value theorem from real numbers to
vectors.

\subsection{LaSalle's invariance principle}
\label{sec:lasalle}

In general, a Lyapunov function does not allow one to conclude
asymptotic convergence when its decrease along solutions is not
strict.
\newterm{LaSalle's invariance principle} (or LaSalle's principle, for
short) is a refinement of Lyapunov's stability theorem to characterize
the asymptotic behavior of solutions.
It shows that the solutions converge to an invariant set where the
derivative of the Lyapunov function ``vanishes'', i.e., every cluster
point belongs to the set.
This has been formalized by Cohen and Rouhling~\cite{cohen2017itp} and
we reuse their development with small adjustments (including updates
to compile with the latest versions of \analysis{}).
 
For a function $f(p,t)$, a set $A$ is called \newterm{invariant} if
for all $p\in A$, $f(p,t')$ remains in $A$ for all future times
\footnote{This definition of invariant is often referred to as
\newterm{forward invariant}.}:
\begin{minted}{ssr}
Definition is_invariant {R} {U : normedModType R}
    (f : U -> R -> U) (A : set U) :=
  forall p, A p -> forall t, 0 <= t -> A (f p t).
\end{minted}

The formalization of LaSalle's invariance principle by
Cohen and Rouhling essentially consists of two theorems about
\newterm{$\omega$-limit sets}.
The $\omega$-limit set for a set \rocqin{A} (\rocqin{limS A},
\cite[\sect{4}]{cohen2017itp}) is defined as the union of all cluster
points of solutions starting at points in \rocqin{A}.
Intuitively, \rocqin{limS A} is the set of points $x$ such that there
is an initial condition $p \in\;$\rocqin{A} and a solution starting at
$p$ comes arbitrarily close to $x$ infinitely often at arbitrarily
large times.

The first theorem is the lemma \rocqin{cvg_to_limS}, which states that
for a compact invariant set \rocqin{A}, every function starting in
\rocqin{A} converges to the $\omega$-limit set \rocqin{limS A}.
\begin{minted}{ssr}
Theorem cvg_to_limS (A : set U) : compact A ->
  is_invariant A ->
  forall p, A p -> sol p @ +oo --> (limS A : set U).
\end{minted}
This lemma does not directly depends on an ODE, though the properties
of the solutions of ODEs will be used in practice to prove that a set
is invariant.

The second theorem depends on the notion of ODE
(\sect{\ref{sec:formal_diff_equa}}).  The formalization of Cohen and
Rouhling uses a slightly different notion for ODEs in which a global
variant of the Cauchy-Lipschitz theorem is actually axiomatized as an
hypothesis (at the time, the authors could not rely on an existing
formalization, at least in \rocq{}).
Before stating the theorem, we explain its hypotheses (here,
\rocqin{U} is a normed module):
\begin{minted}[numbers=left,xleftmargin=3.0ex,escapeinside=88]{ssr}
Record hypos := mk {
  K : set U ; 8\label{line:K}8
  K_compact : compact K ; 8\label{line:Kcompact}8
  sol : U -> R -> U ; 8\label{line:sol}8
  sol0 : forall p, sol p 0 = p ; 8\label{line:sol0}8
  solP_sol : solP K sol ; 8\label{line:solP_sol}8
  sol_cont : forall t,
    {within K, continuous (sol^~ t)} 8\label{line:sol_cont}8 }.
\end{minted}
We assume a set \rocqin{K} (line~\ref{line:K}) which is compact
(line~\ref{line:Kcompact}).
We consider a flow of the (autonomous) system instead of a single
solution (line \ref{line:sol}), i.e., \rocqin{sol} gives a solution for
every initial state $p$.
\rocqin{sol0} (line~\ref{line:sol0}) states the initial condition.
\rocqin{sol} is asked to be continuous w.r.t.\ to the
initial state \rocqin{p} (line~\ref{line:sol_cont}) and required to be a genuine
solution for each \rocqin{p}, i.e., to satisfy the ODE (line \ref{line:solP_sol}).
Even though we are only interested in non-negative times, Cohen and
Rouhling actually also fix values for negative times
\cite[\sect{3.2}]{cohen2017itp} \cite[\sect{5.3}]{rouhling2018cpp}.
More precisely, solutions are asked to satisfy the ODE for
non-negative time (line \ref{line:nonnegative_times}) and to be
symmetric w.r.t.\ the initial state for negative time (line
\ref{line:negative_times}):
\begin{minted}[numbers=left,xleftmargin=3.0ex,escapeinside=88]{ssr}
Definition shifted_odd (y : R -> U) :=
 forall t, t < 0 -> y t = 2 *: y 0 - y (- t). 8\label{line:negative_times}8
Definition is_sol (y : R -> U) := shifted_odd y /\
 (forall t, 0 <= t -> is_derive t 1 y (phi (y t))). 8\label{line:nonnegative_times}8
Definition solP K sol :=
 forall y : R -> U, K (y 0) ->
  is_sol y <-> y = sol (y 0).
\end{minted}
However, since \rocqin{is_derive} at $t=0$ requires a two-sided derivative, 
this extension must be compatible with the right derivative at zero.
Thus, the function is total and differentiable on the
whole set. 

Compared to Cohen and Rouhling, using our formalization we can
discharge the two hypothesis \rocqin{solP_sol} and \rocqin{sol_cont}
by the global variant of the Cauchy-Lipschitz theorem and continuous
dependence shown in \sect{\ref{sec:global_cauchy_lipschitz}} and
\sect{\ref{sec:continuity_initial}}.  The concrete steps for this
construction are given for the tilt example later in
\sect{\ref{sec:convergence_equilibrium_points}}.

The theorem \rocqin{stable_limS} says that given a Lyapunov
function~$V$ which is continuous within \rocqin{K} and its derivative
along solutions is non-positive, then \rocqin{limS K} is contained in
the set of points where the derivative of $V$ along the flow is zero.
\begin{minted}{ssr}
Hypothesis Kinvar : is_invariant sol K.
Lemma stable_limS (V : U -> R) :
  {within K, continuous V} ->
  (forall p t, K p -> 0 <= t ->
    derivable (V \o sol p) t 1) ->
  (forall p, K p -> (V \o sol p)^`() 0 <= 0) ->
  limS K `<=` [set p | (V \o sol p)^`() 0 = 0].
\end{minted}
Note the second theorem depends on the set \rocqin{K} is invariant.

LaSalle’s invariance principle is obtained by combining the two
theorems: \rocqin{stable_limS} restricts the $\omega$-limit set to the
region where the Lyapunov derivative vanishes and \rocqin{cvg_to_limS}
ensures that trajectories starting in a compact invariant set converge
to this limit set.
It should be stressed that LaSalle's principle does not imply that
solutions converge to a single point, but only that all cluster points
lie in the invariant.

\section{Application to the estimation error of the tilt}
\label{sec:application_tilt}

We now apply the developments from the previous sections to the
formalization of \autoref{thm:theorem1} which, regarding the ODE
Equation \eqref{eqn14} (\rocqin{Tilt.eqn}, \sect{\ref{sec:diff_equa_error}}),
(i)~characterizes the reachable set
and
(ii)~establishes the convergence of the solutions towards two
equilibrium points.

Showing that the points $(0,0)$ (\rocqin{point1}) and $(0,2e_z)$
(\rocqin{point2}) are equilibrium points
(\sect{\ref{sec:equilibrium_point}}) essentially amounts to proving that
two constant functions satisfy \rocqin{Tilt.eqn}. This is a matter of
symbolic computation using existing lemmas from \mathcomp{}'s algebra
and \robotrocq{} (e.g., its theory for the dot product and rotations) and using a
new theory of time-varying matrices that extends \analysis{}.

Using the same tools, we also show that the reachable set
(\sect{\ref{sec:equilibrium_point}}) is included in $\Upsilon_1$.
We define $\Upsilon_1$ using the \rocqin{Right} function from
\sect{\ref{sec:tilt_ode}} and the Euclidean norm \rocqin{`| ... |_e}:
\begin{minted}{ssr}
Definition Upsilon1 :=
  [set x : 'rV[R]_6 | `| 'e_2 - Right x |_e = 1].
\end{minted}
To show inclusion in $\Upsilon_1$, we assume that the reachable set is
generated from $\Upsilon_1$ as the initial state:
\begin{minted}{ssr}
Let phi := Tilt.eqn alpha1 gamma.
Lemma tilt_reachable_setS :
  reachable_set phi Tilt.Upsilon1 `<=` Tilt.Upsilon1.
\end{minted}
This allows us to restrain ourselves to $\Upsilon_1$ when studying
\rocqin{Tilt.eqn}.
We also prove that the system is locally Lipschitz and
global existence is then obtained  by proving that trajectories
remain in compact Lyapunov sublevel sets and applying
\rocqin{compact_is_sol_cauchy_infty}.

\subsection{Lyapunov stability for the tilt estimation}
\label{sec:apply_lyapunov}

To prove the stability of the equilibrium point $(0,0)$, Benallegue
and al.\ propose the following Lyapunov function:
$V_1 \mydef \frac{||z_{1}||^2}{2\alpha_1} + \frac{||z_2||^2}{2\gamma}$
($\alpha_1$ and $\gamma$ have been introduced by Equation \eqref{eqn12}).
Its formalization is direct:
\begin{minted}{ssr}
Definition V1 (z1_z2 : 'rV[R]_6) : R :=
  let z1 := Left z1_z2 in
  let z2 := Right z1_z2 in
  `|z1|_e ^+ 2 / (2 * alpha1) + 
      `|z2|_e ^+ 2 / (2 * gamma).
\end{minted}

Proving that \rocqin{V1} is a Lyapunov candidate function is fairly
straightforward as it is just a few lines of scripts of symbolic
manipulations \cite[Lemma \rocqin{V1_is_Lyapunov_candidate}]{code}.
Proving that the derivative of \rocqin{V1} is non-positive requires a
trick. Benallegue et al.\ provide the following upper-bound: 
\[
\dot{V}_1 \leq - \underbrace{\begin{bmatrix} ||z_{1}|| & ||z_2 S(e_z)|| \end{bmatrix}}_{u_1}
                 \underbrace{\begin{bmatrix} 1 & -1/2 \\ -1/2 & 1 \end{bmatrix}}_{u_2}
                 \begin{bmatrix} ||z_{1}|| \\ ||z_2 S(e_z)|| \end{bmatrix} \leq 0. \label{eqn:encadrement}
\]

We observe a quadratic form expression $u_1 u_2 \trmx{u_1}$.
We can therefore exploit a property of quadratic forms:
\begin{minted}{ssr}
Lemma u2_quadratic_form_gt0 (v : 'rV_2) : v != 0 ->
  0 < (v *m u2 *m v^T) 0 0.
\end{minted}
We then conclude on the sign of the derivative of $V_1$:
\begin{minted}{ssr}
Lemma derive_along_V1_le0 (D : R) (f : R -> 'rV_6) :
  f 0 \in Tilt.Upsilon1 ->
  is_sol_cauchy_oo (fun=> phi) 0 D (f 0) f ->
  {in `]0, D[, forall t, differentiable f t} ->
  {in `]0, D[, forall t,
    'D~(f) (Tilt.V1 alpha1 gamma) t <= 0}.
\end{minted}
The corresponding proof scripts are about 130 lines of code of
symbolic manipulations, including the computation of the derivative
of~\rocqin{V1} \cite[Lemma \rocqin{derive_along_V1}]{code}, the proof
of the bound (Lemma \rocqin{V1dot_ub}), and the use of the property of
the quadratic form stated just above.

Finally, we apply Lyapunov's stability theory
(\sect{\ref{sec:lyapunov_stability_thm}}) to conclude that the
equilibrium point \rocqin{point1} (\sect{\ref{sec:equilibrium_point}})
is stable:
\begin{minted}{ssr}
Lemma equilibrium_zero_stable :
  Tilt.point1 \in Init -> Init `<=` Tilt.Upsilon1 ->
  is_stable_at phi Init Tilt.point1.
\end{minted}
Here, \rocqin{Init} is a
set of admissible initial states containing \rocqin{point1} and we additionally 
assume that \rocqin{Init} is contained in \rocqin{Tilt.Upsilon1}.
Intuitively, this result shows that the tilt estimation error
(\sect{\ref{sec:diff_equa_error}}) stays around 0 when starting close
to~0.

\subsection{Convergence to equilibrium points}
\label{sec:convergence_equilibrium_points}

In many applications of LaSalle's principle, the Lyapunov function
is not only used for the sign of its derivative, but also to construct
the compact, invariant regions of the reachable set \cite[\sect{4.2},
p.~128]{khalil2002}.
That is, given a Lyapunov function $V: U \to \mathbb{R}$ and a
constant $c \in \mathbb{R}$, one defines the \newterm{Lyapunov sublevel set}
$\Omega_c \mydef \{ x \in U \, \mid \, V(x) \leq c \}$:
\begin{minted}{ssr}
Definition sublevel (V : U -> R) c :=
  [set x : U | V x <= c].
\end{minted}
Since the Lyapunov function is non-increasing along solutions, every
solution starting in the sublevel set corresponding to its initial
value remains confined in this set for all future times.
When $\Omega_c$ is compact, this provides the setting where LaSalle's
invariance principle can be applied to characterize the asymptotic
behavior.

\smallskip

As far as our application is concerned, our goal is to show that solutions starting
in \rocqin{Tilt.Upsilon1} converge to one of the two equilibrium points
(\rocqin{point1} and \rocqin{point2}).
%
%
For that purpose, we apply LaSalle's invariance principle
(\sect{\ref{sec:lasalle}}).
Due to the way the Lyapunov function $V_1$
(\sect{\ref{sec:apply_lyapunov}}) is defined, the Lyapunov sublevel
sets are already compact for any state \rocqin{p}:
\begin{minted}{ssr}
Definition sublevelV1 (p : U) :=
  sublevel (Tilt.V1 alpha1 gamma)
           (Tilt.V1 alpha1 gamma p).
Lemma compact_sublevelV1 p : compact (sublevelV1 p).
\end{minted}
We further define for each \rocqin{p}, the set \rocqin{sublevelV1Upsilon1 p} as the 
intersection of the sublevel set with \rocqin{Tilt.Upsilon1}.
As \rocqin{Tilt.Upsilon1} is closed (\sect{\ref{sec:application_tilt}}), we get a family 
of compact invariant sets parameterized by the initial value~\rocqin{p}.
Let us first show how to discharge the two hypotheses about the ODE
solution in the LaSalle formalization.  We prove the hypothesis
\rocqin{solP_sol} (\sect{\ref{sec:lasalle}}) by using the global
extension of the Cauchy-Lipschitz theorem developed in
\sect{\ref{sec:global_cauchy_lipschitz}} to construct a forward
solution \rocqin{sol0} for every \rocqin{y0} in \rocqin{Tilt.Upsilon1}
\cite[Lemma \rocqin{tilt_solP'}]{code} and then extend it to
negative time from the LaSalle formalization \cite[Lemma
  \rocqin{tilt_solP}]{code} similarly to the \rocqin{shifted_odd}
predicate from~\sect{\ref{sec:lasalle}}:
\begin{minted}{ssr}
pose sol y0 t := if t < 0 then
  2 *: y0 - sol0 y0 (- t) else sol0 y0 t.
\end{minted}
We can apply the global Cauchy-Lipschitz theorem here since
 \rocqin{sublevelV1Upsilon1 y0} is compact and the solution is 
 contained in this set due to \rocqin{V1} being non-increasing.

The remaining assumption (corresponding to
\rocqin{sol_cont}---\sect{\ref{sec:lasalle}}) is that the flow is
continuous in the initial values, which can be discharged using the
theory developed in \sect{\ref{sec:continuity_initial}}.  The theorem
\rocqin{continuous_dependence} gives continuity of the flow for
positive time, \rocqin{sol p 0 = p} gives it at zero, and for negative
time it follows from the formula above.

By applying \rocqin{stable_limS} (\sect{\ref{sec:lasalle}}), and using
the fact that the only points in \rocqin{Upsilon1} for which
$\dot{V}_1$ is zero are \rocqin{point1} $(0,0)$ and \rocqin{point2}
$(0,2e_2)$ \cite[file \rocqin{tilt_lasalle.v}]{code}, we get:
\begin{minted}{ssr}
Lemma tilt_limS_points p :
  p \in Tilt.Upsilon1 ->
    limS tilt_sol (sublevelV1Upsilon1 p)
    `<=` Tilt.points.
\end{minted}
Here, \rocqin{`<=`} is a notation for subset,
and \rocqin{points} is the two-point set containing \rocqin{point1} and
\rocqin{point2} (\sect{\ref{sec:equilibrium_point}}).
It follows:
\begin{minted}{ssr}
Lemma tilt_cluster_points p : p \in Tilt.Upsilon1 ->
 cluster (tilt_sol p t @[t --> +oo]) `<=` Tilt.points.
\end{minted}
That is, the set of cluster points for any solution starting in
\rocqin{Tilt.Upsilon1} is contained in the two-point set.
However, as the solution is continuous, and only takes values in the
compact set \rocqin{sublevelV1Upsilon1 p}, it follows that its cluster set is
connected \cite[Lemma \rocqin{compact_connected_cluster}]{code}.  Thus, the only
possibilities are the one-point sets or the empty set.
We can further exclude the empty set as every sequence in a compact set has a cluster point,
which can be proven immediately from the definition of compactness in \analysis{}. 
 In summary, we can conclude convergence to one of the two equilibrium points:
\begin{minted}{ssr}
Lemma tilt_cvg_point1_point2 p :
  p \in Tilt.Upsilon1 ->
  (tilt_sol p t @[t --> +oo] --> Tilt.point1) \/
  (tilt_sol p t @[t --> +oo] --> Tilt.point2).
\end{minted}

\section{Related work}
\label{sec:related_work}

ODEs and their applications are a classical topic in mathematics, and
consequently have been the subject of formalization in proof
assistants. In \sect{\ref{sec:formal_diff_equa}}, we presented a
formalization of the local Cauchy-Lipschitz theorem, proving existence
and uniqueness of local solutions of Cauchy problems, as well as
global variants.
Most related work in the \rocq{} prover is focused on constructive or
computational aspects.  Makarov and Spitters
\cite[Lemma~1]{makarov2013itp} formalize the Picard operator and show
that it is a contraction, but the main purpose is to apply the
iteration for exact real computation.  Their formalization is carried
out in a constructive setting and they do not aim to fully relate it
to the classical theory of existence and uniqueness.  Similarly, Thies
\cite{holger2026cpp} proposes a constructive formalization of solution
operators for analytic ODEs used to compute solution trajectories in
the sense of exact real computation.  The method is based on power
series expansions instead of Picard iteration.  Our formalization is
the first textbook-style formalization of the classical
Cauchy-Lipschitz theorems in \rocq{}.

One can also find a formalization of the local Cauchy-Lipschitz
theorem in mathlib but no global variants.
Compared to mathlib, which relies on a kernel level axiom for the 
soundness of quotients, our proof relies only on \mathcomp{}'s 
library-level construction, demonstrating that such an axiom can 
be avoided in practice.
Significant work on the theory of ODEs is already available in
Isabelle/HOL, including the local Cauchy-Lipschitz theorem, the
maximal existence interval, differentiable dependence, and
applications to verified numerical
methods~\cite{immler2012itp,immler2018jar}.
Our formalization does not provide fundamentally new results that are
not available in these developments compared to Isabelle/HOL, but we
believe that the \rocq{} specific aspects such as the use of quotient
types and the integration within the \mathcomp{} ecosystem can be a
useful reference for future design choices.
On the other hand, to the best of our knowledge, neither mathlib nor
Isabelle/HOL contains a general formalization of Lyapunov stability
theory, as we develop it in the paper.  Moreover, applications to
robotics on top of a formalized stability theory appear to have not
yet been explored in Lean or Isabelle/HOL.

The library \robotrocq{} was originally developed to formalize the
forward kinematics problem of robot manipulators and was therefore
limited to statics \cite{affeldt2017cpp}. We had to extend this work
to support our application to humanoid robots by adding analytic
theories such as time-dependent matrices, etc.
There are several pieces of work about formalization of robotics in
HOL4.  For example, Shi et al.\ also develop a theory of derivation of
functional matrices in HOL4 to analyze the Jacobian matrices in the
direct kinematics problem~\cite{shi2018fac}; they apply it to the
Stanford manipulator.  This is a standard problem in robotics but
involving different theories compared to stability. Moreover, HOL4
does not seem to feature a formal theory of ODEs.

Cohen and Rouhling formalized LaSalle's invariance principle in
\rocq{} \cite{cohen2017itp} and applied it to the two-dimensional
problem of the inverted pendulum \cite{rouhling2018cpp}. Our work
reuses their result in a three-dimensional context and we are now able
to discharge the hypotheses they had to introduce to axiomatize the
theory of ODEs. This requires our global variant of the
Cauchy-Lipschitz theorem and continuous dependence.

Other formalizations in proof assistants of Lyapunov stability exist:
Ricketts and al. formalize an equivalent of the Lyapunov's second
theorem with modalities in \rocq{} \cite{ricketts2016emsoft} and prove
the stability of a PID controller \cite{chan2016coqpl}. Devadze and
al. formalize Lyapunov's stability with \minlog{} for discrete-time
polynomial systems \cite{devadze2025jar}. In comparison, we try to
directly formalize the standard framework familiar to roboticians,
with continuous-time differential equations.

\section{Conclusions}
\label{sec:conclusion}

In this paper, we reported on the formalization in the \rocq{} prover
of the convergence analysis of the error dynamics of a tilt estimator,
a key issue in the design of humanoid robots.
We started by extending an existing library for verification of static
aspects of robot manipulators (namely, \robotrocq{}) with velocity
(using the \analysis{} library). This made it possible to formally
model the physical problem of tilt estimation
(\sect{\ref{sec:formal_model}}). As the foundation of a theory of
ODEs, we formalized the Cauchy-Lipschitz theorem, continuation to
global solutions, and continuous dependence on initial conditions.  We
also discussed the difficulties stemming from the correct
specification of boundary conditions
(\sect{\ref{sec:formal_diff_equa}}). Based on these extensions, we
formalized stability theory up to Lyapunov's stability theorem
(\sect{\ref{sec:stability_theory}}), that we used to reproduce the
arguments of Benallegue et al., in particular the formalization of a
Lyapunov function that proves Lyapunov stability, and that can be
further used to prove convergence towards equilibrium points (using
LaSalle's invariance principle that we adapted from existing work)
(\sect{\ref{sec:application_tilt}}).
A quantitative overview of the complete accompanying \rocq{}
development is provided in \autoref{app:formalization}.

We believe that our experiment is valuable because we discuss
formalization of mathematical theories that otherwise often lack
details when couched on paper: the boundary conditions of even basic
results about ODEs are easily glossed over and the relation between
stability theory and ODEs is often implicit in textbook presentations.
More concretely, our theory of ODEs fills a pressing need in the
\rocq{} ecosystem that has so far been lacking a general-purpose formalization
of the Cauchy-Lipschitz theorem for applications to continuous-dynamics verification
(see, e.g., \cite[p.~282:20]{daggitt2026icfp}).

As explained in \sect{\ref{sec:formal_model}}, the present
formalization proves the first claim of
\cite[Theorem~1]{benallegue2023tac}: the set $\Upsilon_1$ is invariant, the
system has exactly two equilibrium points, and every solution starting
in $\Upsilon_1$ converges to one of them.  The remaining claims are
that the origin is almost globally asymptocially stable, and that the
parameters can be adjusted to make the convergence arbitrarily fast in
any fixed compact set that excludes the other equilibrium point.
Formalizing these claims requires a number of additional results, e.g.,
on exponential stability, linearization theory, and unstable manifolds.
In future work, we plan to extend our formalization in these directions.

\begin{acks}
The authors are grateful to Mehdi Benallegue for his guidance and
acknowledge partial support from Inria's French-Japanese Inria
Associated Team FormaSys.
\end{acks}

\bibliography{tilt_ode}
\appendix
\section{Organization of the Rocq Development}
\label{app:formalization}
The accompanying Rocq development is organized into the files listed
in \autoref{tab:files}.
The instructions for compiling can be found in \rocqin{README.md}.
\begin{table*}
\begin{tabular}{l|l|r}
file & contents & l.o.c.\\
\hline
\rocqin{tilt_analysis.v} & additions to \analysis{} \cite{analysis} & 969 \\
\rocqin{gronwall.v} & Gronwall-Bellman inequality \cite[Lemma A.1]{khalil2002}, used in \sect{\ref{sec:continuity_initial}} & 409 \\
\rocqin{vector_integral.v} & component-wise integration (\sect{\ref{sec:about_formal_proof}}) & 95 \\
\rocqin{tilt_robot.v} & additions to \robotrocq{} \cite{affeldt2017cpp} & 423 \\
\rocqin{lasalle.v} & LaSalle's invariance principle adapted from \cite{cohen2017itp} (\sect{\ref{sec:lasalle}}) & 517 \\
\rocqin{ode_common.v} & preliminaries for the formalization of ODEs & 640 \\
\rocqin{ode_contseg.v} & quotient of continuous functions over an interval (\sect{\ref{sec:quot_contSeg}}) & 788 \\
\rocqin{picard_contraction.v} & proof that the Picard operator is a contraction (discussed throughout \sect{\ref{sec:formal_diff_equa}}) & 819 \\
\rocqin{ode_local.v} & local Cauchy-Lipschitz theorem (\sect{\ref{sec:fwd_cauchy_lipschitz_proof}}, \sect{\ref{sec:gen_sym_case}}) & 1457 \\
\rocqin{ode_global.v} & global existence (\sect{\ref{sec:global_cauchy_lipschitz}}) and continuous dependency on initial conditions (\sect{\ref{sec:continuity_initial}}) & 1326 \\
\rocqin{tilt_stability.v} & stability theory (\sect{\ref{sec:equilibrium_point}}, \sect{\ref{sec:lyapunov_functions}}), including Lyapunov's stability theorem (\sect{\ref{sec:lyapunov_stability_thm}}) & 920 \\
\rocqin{tilt_lyapunov.v} & physical model (\sect{\ref{sec:formal_model}}), & \\
                         & application of \rocqin{tilt_stability.v} to \autoref{thm:theorem1} (Lyapunov stability, \sect{\ref{sec:apply_lyapunov}}) & 1278 \\
\rocqin{tilt_lasalle.v} & application of \rocqin{lasalle.v} to \autoref{thm:theorem1} (convergence to equilibrium points, \sect{\ref{sec:convergence_equilibrium_points}}) & 873 \\
\hline
\multicolumn{2}{r|}{Total (excluding \rocqin{lasalle.v})} & 9997 \\
\end{tabular}
\caption{Overview of the accompanying material \cite{code} with references to sections in this paper}
\label{tab:files}
\end{table*}
\end{document}